\documentclass[twocolumn,epjc3-upd,fleqn]{svjour3}
\input{preamble.tex}
\begin{document}
\newcommand{\WIMPAnalysisEnergyRange}{$10-28\1{keV}$}
\newcommand{\AnalysisFirstMonth}{{November 2013}}
\newcommand{\AnalysisLastMonth}{{March 2015}}
\newcommand{\NumberOfChannels}{{26}}
\newcommand{\NumberOfXrayPeaksDoubleCoincidenceChannels}{{25}}
\newcommand{\NumberOfXrayPeaksAntiCoincidenceChannels}{one}
\newcommand{\TotalExposure}{{$23.15\1{kg\cdot yr}$}}
\newcommand{\SelectedExposure}{{$20.02\1{kg\cdot yr}$}}
\newcommand{\SelectedThExposure}{{$18.56\1{kg\cdot yr}$}}
\newcommand{\SelectedThExposureAtThreshold}{{$1.6\1{kg\cdot yr}$}}
\newcommand{\WIMPExposure}{\todo{{$13\1{kg\cdot yr}$}}}
\newcommand{\SigmaSIMinimum}{\todo{{$xx \1{cm^{2}}$}}}
\newcommand{\SigmaSIMinimumWIMPMass}{\todo{{$40 \1{GeV}/c^{2}$}}}
\newcommand{\LowEnergyBackgroundSpectrumMax}{$60 \1{keV}$}
\newcommand{\CUORETotalMass}{$742\1{kg}$}
\newcommand{\CUOREOperatingTemperature}{$\sim 10\1{mK}$}
\newcommand{\QinoLimit}{$2.8 \times 10^{24}\1{yr}$}
\newcommand{\QZQinoCombinedLimit}{$4.0 \times 10^{24}\1{yr}$}
\newcommand{\CUOREResolution}{$\sim 5\1{keV}$}
\newcommand{\TeQvalue}{$2527.5\1{keV}$}
\newcommand{\TrigThresholdLower}{30}
\newcommand{\TrigThresholdUpper}{$120\1{keV}$}
\newcommand{\CCVREnergyThreshold}{$3\1{keV}$}
\newcommand{\CCVRTriggerEfficiency}{$> 90\%$}
\newcommand{\QZActiveMass}{$38.4\1{kg}$}
\newcommand{\QZInstrumentedMass}{$39.1\1{kg}$}
\newcommand{\QZOperationTemperature}{$\sim 10\1{mK}$}
\newcommand{\QZOTUpper}{$12\1{keV}$}
\newcommand{\NPulserEnergyRangeMin}{0}
\newcommand{\NPulserEnergyRangeMax}{$50\1{keV}$}
\newcommand{\TeXrayPeakRangeMin}{26}
\newcommand{\TeXrayPeakRangeMax}{$32\1{keV}$}
\newcommand{\TeXrayFirstPeakPos}{$27 $}
\newcommand{\TeXraySecondPeakPos}{$31\1{keV}$}
\newcommand{\TeXrayFirstPeakPosRes}{\todo{$0.17 \pm 0.05\1{keV}$}}
\newcommand{\TeXraySecondPeakPosRes}{\todo{$0.24 \pm 0.15\1{keV}$}}
\newcommand{\CalibrationUncertainty}{\todo{$0.18\1{keV}$}}
\newcommand{\MCPeakPositionDiff}{$0.01\1{keV}$}
\newcommand{\NRQuenchingUncertainty}{{$7\%$}}
\newcommand{\LowEnergyResolutionSigma}{$\pm (\sigma_\n{DS}+1\1{keV})$}
\newcommand{\LowEnergyPulserMin}{13}
\newcommand{\LowEnergyPulserMax}{$27\1{keV}$}
\newcommand{\TeXrayFirstPeakPosShiftAllowance}{$0.6\1{keV}$}
\newcommand{\CoincidenceTime}{$\pm 5\1{ms}$}
\newcommand{\DaughterNucleiEnergy}{$\sim 100\1{keV}$}
\newcommand{\LECoincidenceTime}{$\pm 100\1{ms}$}
\newcommand{\ThirtySevenkeVPeakDiff}{$0.03\pm 0.06\1{keV}$}
\newcommand{\FinalEnergyThresholdMin}{8}
\newcommand{\FinalEnergyThresholdMinInSpc}{$10\1{keV}$}
\newcommand{\FinalEnergyThresholdMax}{$35\1{keV}$}
\newcommand{\NumberOfDsBoPairsDiscardedByOTLarger}{37}
\newcommand{\NumberOfDsBoPairsWithThLower}{16}
\newcommand{\NumberOfDsBoPairs}{490}
\newcommand{\TotalNumberOfDsBoPairs}{539}
\newcommand{\BkgAtFiftykeV}{$0.05\1{counts/(kg \cdot keV \cdot day)}$}
\newcommand{\BkgAtTenkeV}{$1.7\1{counts/(kg \cdot keV \cdot day)}$}

\newcommand{\NumberOfDsChPairs}{\todo{508}}
\newcommand{\OTChisquareValueObtainmentEnergyRegion}{$35 - 50\1{keV}$}
\newcommand{\AntiCoincidenceEff}{$99.2 \pm 0.3\%$}
\newcommand{\TotalEff}{\todo{$89.3 \pm xx\%$}}
\newcommand{\LowEPeakLow}{31}
\newcommand{\LowEPeakLowkeV}{$31\1{keV}$}
\newcommand{\LowEPeakHigh}{$37\1{keV}$}
\newcommand{\OTCutUncertaintyMin}{$0.3$}
\newcommand{\OTCutUncertaintyMax}{$1\%$}
\newcommand{\CalibrationPeakFitRangeMin}{22}
\newcommand{\CalibrationPeakFitRangeMax}{$34\1{keV}$}
\newcommand{\EffCorRateCompZeroEne}{$\sim 25\1{keV}$}
\newcommand{\StandCalibUncertaintyAtROI}{$0.05 \pm 0.05\,(\n{stat.})\pm 0.09\,(\n{syst.}) \1{keV}$}
\newcommand{\MostIntenseLowEPeakCal}{$27.60 \pm 0.05 \1{keV}$}
\newcommand{\MostIntenseLowEPeakBkg}{$27.65 \pm 0.13 \1{keV}$}
\newcommand{\MostIntenseLowEPeakDiffCal}{\mbox{$0.13 \pm 0.05 \1{keV}$}}
\newcommand{\MostIntenseLowEPeakDiffBkg}{$0.18 \pm 0.13 \1{keV}$}
\newcommand{\KSProbThr}{0.1}

\newcommand{\WIMPLocalDensity}{$\rho_{\chi} = 0.3\1{GeV}/{c}^2$}
\newcommand{\LocalCircularVelocity}{$v_0 = 220\1{km/s}$}
\newcommand{\EarthOrbitalVelocity}{$v_{\n{orb}} = 29.8\1{km/s}$}
\newcommand{\GalacticEscapeVelocity}{$v_{\n{esc}}=650\1{km/s}$}
\newcommand{\VescImpactUpLimit}{$10^{-5}\n{pb}$}


\title{Low Energy Analysis Techniques for CUORE}

\author{
C.~Alduino\thanksref{USC} 
\and
K.~Alfonso\thanksref{UCLA} 
\and
D.~R.~Artusa\thanksref{USC,LNGS} 
\and
F.~T.~Avignone~III\thanksref{USC} 
\and
O.~Azzolini\thanksref{INFNLegnaro} 
\and
G.~Bari\thanksref{INFNBologna} 
\and
J.W.~Beeman\thanksref{LBNLMatSci} 
\and
F.~Bellini\thanksref{Roma,INFNRoma} 
\and
G.~Benato\thanksref{BerkeleyPhys} 
\and
A.~Bersani\thanksref{INFNGenova} 
\and
M.~Biassoni\thanksref{Milano,INFNMiB} 
\and
A.~Branca\thanksref{INFNPadova} 
\and
C.~Brofferio\thanksref{Milano,INFNMiB} 
\and
C.~Bucci\thanksref{LNGS} 
\and
A.~Camacho\thanksref{INFNLegnaro} 
\and
A.~Caminata\thanksref{INFNGenova} 
\and
L.~Canonica\thanksref{MIT,LNGS} 
\and
X.~G.~Cao\thanksref{Shanghai} 
\and
S.~Capelli\thanksref{Milano,INFNMiB} 
\and
L.~Cappelli\thanksref{LNGS} 
\and
L.~Cardani\thanksref{INFNRoma} 
\and
P.~Carniti\thanksref{Milano,INFNMiB} 
\and
N.~Casali\thanksref{INFNRoma} 
\and
L.~Cassina\thanksref{Milano,INFNMiB} 
\and
D.~Chiesa\thanksref{Milano,INFNMiB} 
\and
N.~Chott\thanksref{USC} 
\and
M.~Clemenza\thanksref{Milano,INFNMiB} 
\and
S.~Copello\thanksref{Genova,INFNGenova} 
\and
C.~Cosmelli\thanksref{Roma,INFNRoma} 
\and
O.~Cremonesi\thanksref{INFNMiB} 
\and
R.~J.~Creswick\thanksref{USC} 
\and
J.~S.~Cushman\thanksref{Yale}
% ...
} 

\institute{
Department of Physics and Astronomy, University of South Carolina, Columbia, SC 29208, USA\label{USC} 
\and
Department of Physics and Astronomy, University of California, Los Angeles, CA 90095, USA\label{UCLA} 
\and
INFN -- Laboratori Nazionali del Gran Sasso, Assergi (L'Aquila) I-67100, Italy\label{LNGS} 
\and
INFN -- Laboratori Nazionali di Legnaro, Legnaro (Padova) I-35020, Italy\label{INFNLegnaro} 
\and
INFN -- Sezione di Bologna, Bologna I-40127, Italy\label{INFNBologna} 
\and
Materials Science Division, Lawrence Berkeley National Laboratory, Berkeley, CA 94720, USA\label{LBNLMatSci} 
\and
Dipartimento di Fisica, Sapienza Universit\`{a} di Roma, Roma I-00185, Italy\label{Roma} 
\and
INFN -- Sezione di Roma, Roma I-00185, Italy\label{INFNRoma} 
\and
Department of Physics, University of California, Berkeley, CA 94720, USA\label{BerkeleyPhys} 
\and
INFN -- Sezione di Genova, Genova I-16146, Italy\label{INFNGenova} 
\and
Dipartimento di Fisica, Universit\`{a} di Milano-Bicocca, Milano I-20126, Italy\label{Milano} 
\and
INFN -- Sezione di Milano Bicocca, Milano I-20126, Italy\label{INFNMiB} 
\and
INFN -- Sezione di Padova, Padova I-35131, Italy\label{INFNPadova} 
\and
Massachusetts Institute of Technology, Cambridge, MA 02139, USA\label{MIT} 
\and
Shanghai Institute of Applied Physics, Chinese Academy of Sciences, Shanghai 201800, China\label{Shanghai} 
\and
Dipartimento di Fisica, Universit\`{a} di Genova, Genova I-16146, Italy\label{Genova} 
\and
Department of Physics, Yale University, New Haven, CT 06520, USA\label{Yale} 
% ...
} 

\thankstext{fn1}{Deceased}
\thankstext{fn2}{Presently at: INFN -- Laboratori Nazionali di Frascati, Frascati (Roma) I-00044, Italy}

\date{Updated on \today}

\maketitle
\begin{abstract}
CUORE is a tonne-scale cryogenic detector operating at the Laboratori
Nazionali del Gran Sasso (LNGS) that uses tellurium dioxide bolometers to search 
for neutrinoless double-beta decay of {\Te}. 
CUORE is also suitable to search for low energy rare events such as solar axions or 
WIMP scattering, thanks to its ultra-low background and large target mass. 
However, to conduct such sensitive searches requires 
improving the energy threshold to 10 keV.
In this paper, we describe the analysis
techniques developed for the low energy analysis of CUORE-like
detectors, using the data acquired from {\AnalysisFirstMonth} to
{\AnalysisLastMonth} by {\qz}, a single-tower prototype designed to
validate the assembly procedure and new cleaning techniques of CUORE.
We explain the energy threshold optimization, continuous monitoring of
the trigger efficiency, data and event selection, and energy
calibration at low energies in detail. We also present the low energy
background spectrum of {\qz} below
{\LowEnergyBackgroundSpectrumMax}. Finally, we report the sensitivity
of CUORE to WIMP annual modulation using the {\qz} energy threshold
and background, as well as an estimate of the uncertainty on the
nuclear quenching factor from nuclear recoils in {\qz}.
\end{abstract}

\section{Introduction}
CUORE (Cryogenic Underground Observatory for Rare Events) is a
tonne-scale cryogenic detector primarily designed to search for the
neutrinoless double-beta ($0\nu\beta\beta$) decay of
{\Te}~\cite{Arnaboldi:2002du,Artusa:2014lgv}. In {\BBless}, two
neutrons in an atomic nucleus simultaneously decay to two protons and
two electrons, without emitting any electron antineutrinos. 
The experimental signature of {\BBless} is a 
sharp peak at the tail end of the two-neutrino double-beta decay (\TwoNuBB)
summed energy spectrum.
The bolometric technique of CUORE offers an excellent
energy resolution of {\CUOREResolution} FWHM at the $Q$-value of {\Te},
{\TeQvalue}~\cite{Redshaw:2009cf,Scielzo:2009nh,Rahaman2011412}, which
suppresses the $2\nu\beta\beta$ decay background leaking into the
{\BBless} signal region of interest (ROI)~\cite{Alduino:2016vtd}.

The CUORE program builds on a predecesor experiment, Cuoricino, which reported a lower
limit on the {\Te} {\BBless} half-life of {\QinoLimit} (90\% C.L.)
with data accumulated from 2003 to 2008~\cite{Andreotti:2010vj}. The
successor experiment {\qz}, operated from 2013 to 2015, set a limit of
{\QZQinoCombinedLimit} (90\% C.L.) in combination with the {\qino}
data~\cite{Alfonso:2015wka}. CUORE is currently in data-taking at LNGS.

While CUORE will be one of the leading {\BBless} experiments during
its scheduled 5 years of data-taking, it will also benefit from the
ultra-low background and large target mass to search for lower energy rare
events, such as the direct detection of Weakly Interacting Massive
Particle (WIMP) dark matter or solar axions~\cite{Arnaboldi:2003tu}. WIMP direct detection is
possible with terrestrial detectors by measuring nuclear recoils
produced as WIMPs scatter off nuclei in the target
material~\cite{Goodman:1984dc}. The resulting energy spectrum falls
quasi-exponentially as a function of energy and extends to only a few
tens of keV for typical WIMPs with masses of
$\mathcal{O}(100\1{GeV/c^2})$. For WIMPs in the galactic halo, an
annual modulation of the event rate
is expected due to the Earth's motion relative to the dark
matter halo of the Milky Way~\cite{Drukier:1986tm,Freese:1987wu},
with event rates highest in June, when the Earth's relative velocity 
with respect to the halo is maximal, and lowest in December. 
Alternatively, solar axions can be detected by the inverse Primakoff effect in the
Coulomb field of the crystal, with a signal from the M1 transition of
$^{57}\n{Fe}$ expected at
$14.4\1{keV}$~\cite{Alessandria:2012mt}. A critical requirement for
both these searches is the achievement of an energy threshold of $<
10\1{keV}$ and sufficient rejection of low energy noise and/or spurious
events, in addition to a detailed understanding of the low energy backgrounds and adequate detector stability.

Since Cuoricino, which proved that the CUORE detector technology is
well suited for searching for {\BBless}~\cite{Andreotti:2010vj}, the
CUORE collaboration has worked to lower the energy thresholds to
perform low energy rare event searches. We developed a new low energy
software trigger, the ``optimal trigger''
(OT)~\cite{DiDomizio:2010ph}, based on filtering the
continuous data stream before the application of the trigger condition. We
implemented this algorithm in test measurements of the bolometric performance 
of a small number of CUORE crystals in a dedicated setup (CUORE
Crystal Validation Runs, CCVR)~\cite{Alessandria:2011vj}. The results
were encouraging; we were able to identify events with energies as low
as {\CCVREnergyThreshold} with a trigger efficiency of
{\CCVRTriggerEfficiency} in three
bolometers out of four~\cite{Alessandria:2012ha}.
In {\qz}, we improved the OT technique and developed new software and
hardware tools for the low energy analysis. These tools include the
continuous monitoring of the trigger efficiency, the development of
low energy event selection criteria, and a low energy calibration.

In this paper, we describe our low energy analysis techniques in
detail, present the energy thresholds and spectrum from the {\qz}
experiment and report the sensitivity of CUORE to WIMP-induced annual modulation assuming the same energy thresholds
and background. We also evaluate the uncertainty on the nuclear quenching
factor of {\TeO} obtained from {\qz} data. Specifically,
Section~\ref{sec:exp_setup} describes the experimental setup and data
production of {\qz}. Section~\ref{sec:OT} explains the OT algorithm and
trigger efficiency evaluation, and section~\ref{sec:data_selection}
details the data and event selection
criteria. Section~\ref{sec:low_energy_spectrum} presents the low energy
spectrum as well as 
the determination of the analysis threshold and the evaluation of the low energy calibration uncertainty.
Section~\ref{sec:wimp_analysis} outlines the analysis
developed for a WIMP search, including the nuclear quenching factor of
{\TeO} estimation, as well as a study of the sensitivity of CUORE to
WIMP-induced annual modulation.  Finally, we present the summary in
Sec.~\ref{sec:summary}.

\section{The CUORE-0 experiment}
\label{sec:exp_setup}
{\qz} comprised 52 {\TeO} crystals with a total active mass of
{\QZActiveMass}. The crystals were arranged in a single tower, with 13
planes of four $5\times5\times5\1{cm^3}$ crystals held securely inside
a copper frame by polytetrafluoroethylene supports. The detector was
hosted in the same cryostat used for {\qino} 
at a base temperature of {\QZOperationTemperature},
and used the same shielding and
electronics. The detector design, construction, and operation are
detailed in~\cite{Alduino:2016vjd}.  Each crystal was instrumented
with a neutron-transmutation-doped (NTD) germanium thermistor~\cite{Haller1984} to read
the thermal signal, and a silicon resistor~\cite{Andreotti:2012zz}, used as a Joule heater to
inject reference pulses of constant energy every $300\1{s}$.  The
reference pulses were mainly used to correct the thermal gain against
the drift in temperature, but they also played a fundamental role in
determining the OT efficiency, as explained in Sec.~\ref{sec:OT}.
%
The thermal readout of each bolometer thermistor was in the form of a
voltage waveform continuously acquired with a sampling frequency of
125~S/s.  Taking advantage of this relatively low acquisition rate, we
could record the continuous data stream without hardware
trigger. This allowed us to reprocess the raw data with different
software trigger algorithms and to optimize the energy thresholds
offline.

We collected data in one-day-long runs. Approximately once per month, we
calibrated the detector by irradiating it for about 3 days using
thoriated tungsten wires inserted between the outer vacuum chamber of
the cryostat and the external lead shielding. The basic analysis unit
is the dataset, which is composed of initial calibration runs,
approximately 3 weeks of physics runs, and final calibration runs.
For the low energy analysis, we also performed a dedicated measurement
before each final calibration using low energy pulses generated by the
Joule heater, with energies ranging from {\NPulserEnergyRangeMin} to
{\NPulserEnergyRangeMax}.

A comprehensive description of the standard {\qz} data processing
procedure for {\BBless} and {\TwoNuBB} decay can be found
in~\cite{Alduino:2016zrl}. The following summarizes the major steps of
the data processing that are common to both low energy and high energy
({\BBless} and {\TwoNuBB} decay) analyses. After application of the software
trigger, we store events in 5~s windows and evaluate the pulse
amplitude using the optimal filter (OF)~\cite{Gatti:1986cw}. The OF
weights each frequency component by the expected signal-to-noise
ratio, calculated as the ratio between the average pulse (obtained
from the $2615\1{keV}$ $\gamma$ rays in the calibration data), and the
average noise power spectra (NPS). 
To calculate the NPS we average baselines recorded in windows without a signal, 
that are acquired simultaneously on all bolometers every 200~s and selected 
for the NPS after some quality checks (basically we require that no pulses or 
pulse tails are present in the window).
The drift in signal gain due to temperature
fluctuations in the bolometer is corrected by performing a linear
regression between the detector baseline voltage, a proxy for the
detector temperature, and the amplitude of the reference pulser
events. Finally, the voltage readout is converted to energy using the
numerous $\gamma$ ray peaks between 511 and $2615\1{keV}$ from the
daughter nuclei of {\Th} in the calibration data. The mapping from
pulse amplitude to energy is described as a second-order polynomial
with zero intercept to take into account possible nonlinearities, 
as those originated from the pulse shape dependence on energy.

\section{Optimal trigger optimization and efficiency evaluation}
\label{sec:OT}
The energy ROI of the standard data processing
for {\BBless} analysis is in the MeV range, and we use a simple
trigger algorithm which flags an event when the slope of the waveform
exceeds a given threshold for a certain amount of
time~\cite{Alduino:2016vjd}.  This results in energy thresholds
ranging from {\TrigThresholdLower} to {\TrigThresholdUpper}, depending
on the bolometer, while the use of the OT algorithm is critical for
lowering the threshold below $30\1{keV}$.

The OT algorithm works as follows. The data buffer is divided into
slices that are continuously filtered in the frequency domain with the
OF described in Sec.~\ref{sec:exp_setup}.  The filtered waveforms are
less noisy than the original waveforms, and baseline fluctuations are
reduced. This allows us to trigger on the filtered trace in the time
domain with a simple threshold as low as $< 10\1{keV}$. Furthermore,
the filter is sensitive to the shape of the expected signal,
suppressing trigger on spurious noise-induced pulses.

The algorithm used in {\qz} is improved relative to that described
in~\cite{DiDomizio:2010ph} in order to achieve a higher trigger
efficiency. In particular, we have removed the veto around high energy
pulses that prevented the algorithm from re-triggering the symmetric
side lobes generated by the OF, which was identified
in~\cite{DiDomizio:2010ph} as the main source of trigger inefficiency.
The new algorithm recognizes the side lobes of a high energy pulse as
OF artifacts and does not flag them.

We set the trigger threshold independently for every bolometer in
every dataset (hereafter ``BoDs'') based on the noise level.  First,
we calculate an OT trigger level at $\theta=3\sigma_\n{OF}$, where
$\sigma_\n{OF}$ is the baseline resolution after applying the OF. The
energy-dependent trigger efficiency $\epsilon(E)$ is modeled by the
Gaussian cumulative density function
\begin{align} \label{eq:trig_eff}
\epsilon(E) = \frac{1}{2} \erf\left(\frac{E-\theta}{\sqrt 2 \sigma_\textrm{OF}}\right)
  + \frac{1}{2},
\end{align}
which is 50\% for $E=\theta$. At this energy we reject 99.86\% of baseline noise.
The trigger threshold, {\TT}, is set to
the value at which 99\% efficiency is reached.  The validity of the
efficiency calculation is checked at the end of each dataset with
dedicated measurements injecting low energy pulses~\cite{Piperno:2016bna}.

\begin{figure}[htbp]
\centering 
\includegraphics[width=0.5\textwidth]{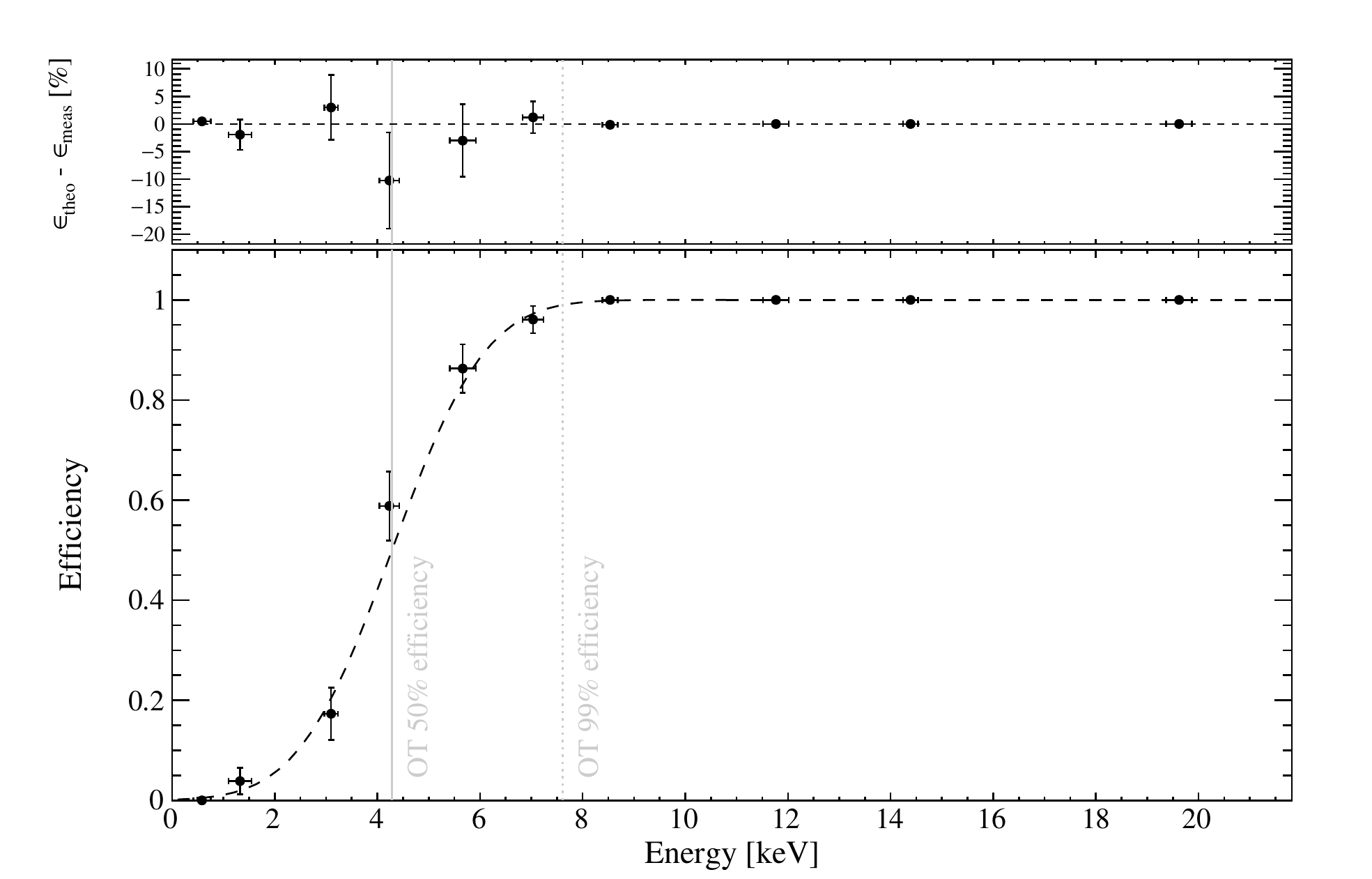}
\caption{(Bottom): An example of the trigger efficiency obtained from
  Eq.~\ref{eq:trig_eff} as a function of energy (dashed line) and data
  obtained from low energy pulser measurements (circles). Energies
  where OT trigger efficiency reaches 50\% ($\theta$) and 99\% ({\TT})
  are shown as vertical gray lines. (Top): The difference between the
  model and the data.}
\label{fig:trig_eff}
\end{figure}
The bottom plot of Fig.~\ref{fig:trig_eff} shows an example of trigger
efficiency (dashed line) as a function of energy, along with data
obtained from corresponding low energy pulser measurements
(circles). Vertical gray lines indicate the energies where OT trigger
efficiency reaches 50\% ($\theta$) and 99\% ({\TT}). In this case,
{\TT} threshold is set at $7.5\1{keV}$. The top plot shows the
difference between the modeled trigger efficiency and the data.
 The {\qz} {\TT} thresholds range from 4
to {\QZOTUpper} for most BoDs, slightly above those of the CCVR
measurement due to a larger noise contribution, as explained in
Section~\ref{subsec:threshold}.
In Table~\ref{tab:thresholdNames} we summarize the different energy thresholds considered in 
this work together with the range of values obtained for the CUORE-0 BoDs.
\begin{table}[htbp]
  \centering
  \caption{Different energy thresholds considered in this work (see Secs~\ref{sec:OT} and \ref{sec:data_selection}). 
The last column represent the range of values obtained for the CUORE-0 BoDs.}
  \label{tab:thresholdNames}
  \begin{tabular*}{1.0\columnwidth}{@{\extracolsep{\fill}} l l l l l}
    \hline\hline Symbol & Name & Description & Energy range (keV) \\
    \hline
$\theta$   & OT trigger level  & \pbox[t][][t]{2.3cm}{Trigger firing \\ energy} & 2 - 7 \\
 {\TT}     & Trigger threshold & \pbox[t][][t]{2.3cm}{Energy for 99\% \\trigger efficiency } & 4 - 12 \\ 
{\Ethr}    & Energy threshold  & \pbox[t][][t]{2.3cm}{Lower noise-free \\energy}            & 8 - 35 \\
    \hline \hline
  \end{tabular*}
\end{table}

\section{Data selection and energy threshold determination}
\label{sec:data_selection}
In this section we detail three steps to select legitimate low energy events.
First, we only choose data whose low energy response and stability are
verified.
Second, we identify and remove non-legitimate events that pass the
trigger requirement, such as electronic noise, tower vibrations,
pile-ups or particle interactions in a thermistor, where the last
appears as a narrow pulse with fast decay time.  Last, we remove
events that occur simultaneously in more than one bolometer since the
probability that WIMP or solar axion interactions occur in more than
one bolometer within the coincidence window is essentially zero.
In this way we reject muons passing through the tower and 
radioactive decays that deposit energy in several bolometers 
($\alpha$-decays in the surfaces of the crystals, 
Compton scattering, cascade $\gamma$-rays...).

\subsection{Dataset-bolometer selection criteria}
\label{subsec:ds_ch_selection_criteria}
While {\qz} ran from March 2013 to March 2015, we only use 11
datasets from the second data-taking campaign, lasting from
{\AnalysisFirstMonth} to {\AnalysisLastMonth},
because of its more stable cryogenic conditions. 
We exclude some runs (a total of $\sim 5$~data-taking days) with an
abnormally higher ($\sim$ 10 times) low energy event rate, which we
attribute to cryostat instability following a helium refill. To
preserve the data quality, we reject the time intervals for each
bolometer that exhibit degraded bolometric performance due to large
baseline excursions or elevated noise levels, as described
in~\cite{Alduino:2016zrl}. The total exposure after these preliminary
quality checks is {\TotalExposure} of {\TeO}.
 
To ensure a stable energy calibration and sufficient resolution at low
energies, we additionally require that the 
residual gain variation of the energy pulser of every BoDs after temperature stabilization
do not vary more than
{\LowEnergyResolutionSigma} from the mean over the entire data-taking
period, where $\sigma_\n{DS}$ is the uncertainty in the pulser
position associated with the dataset and the pulser energy ranges from
{\LowEnergyPulserMin} to {\LowEnergyPulserMax}.  We also discard BoDs
with fewer than 11~events in the region where we
evaluate the pulse shape parameter event selection efficiency
({\OTChisquareValueObtainmentEnergyRegion}, see
Sec.~\ref{subsec:event_selection_criteria}).
%
Finally, we exclude run-bolometer pairs with baseline RMS values that
are greater than $2\sigma$ above the median, where median and $\sigma$
are calculated for each bolometer over all the datasets. After all the
data selection, we use {\NumberOfDsBoPairs} BoDs out of
{\TotalNumberOfDsBoPairs} with a total {\TeO} exposure of
{\SelectedExposure}.

\subsection{Event selection criteria}
\label{subsec:event_selection_criteria}

For the standard {\BBless} analysis we use a set of six pulse shape
parameters to select physical events in {\TeO} based on the pulse shape
characteristics~\cite{Alduino:2016zrl}. These parameters, however, 
lose rejection power at low energy due to the worse signal-to-noise ratio. 
Instead, the OT algorithm provides us
with a powerful shape parameter, {\OT}, to select legitimate signal
candidate events at low energy~\cite{DiDomizio:2010ph}.  We define
{\OT} as the reduced ${\chi}^{2}$ computed between the triggered event
and a cubic spline of the filtered average pulse obtained from the
calibration $\gamma$ rays at $2615 \1{keV}$.
This parameter is sensitive to the shape of the expected signal, 
suppressing pulses with shape deviating from the nominal one.

\begin{figure}[htbp]
\centering 
\includegraphics[width=0.5\textwidth]{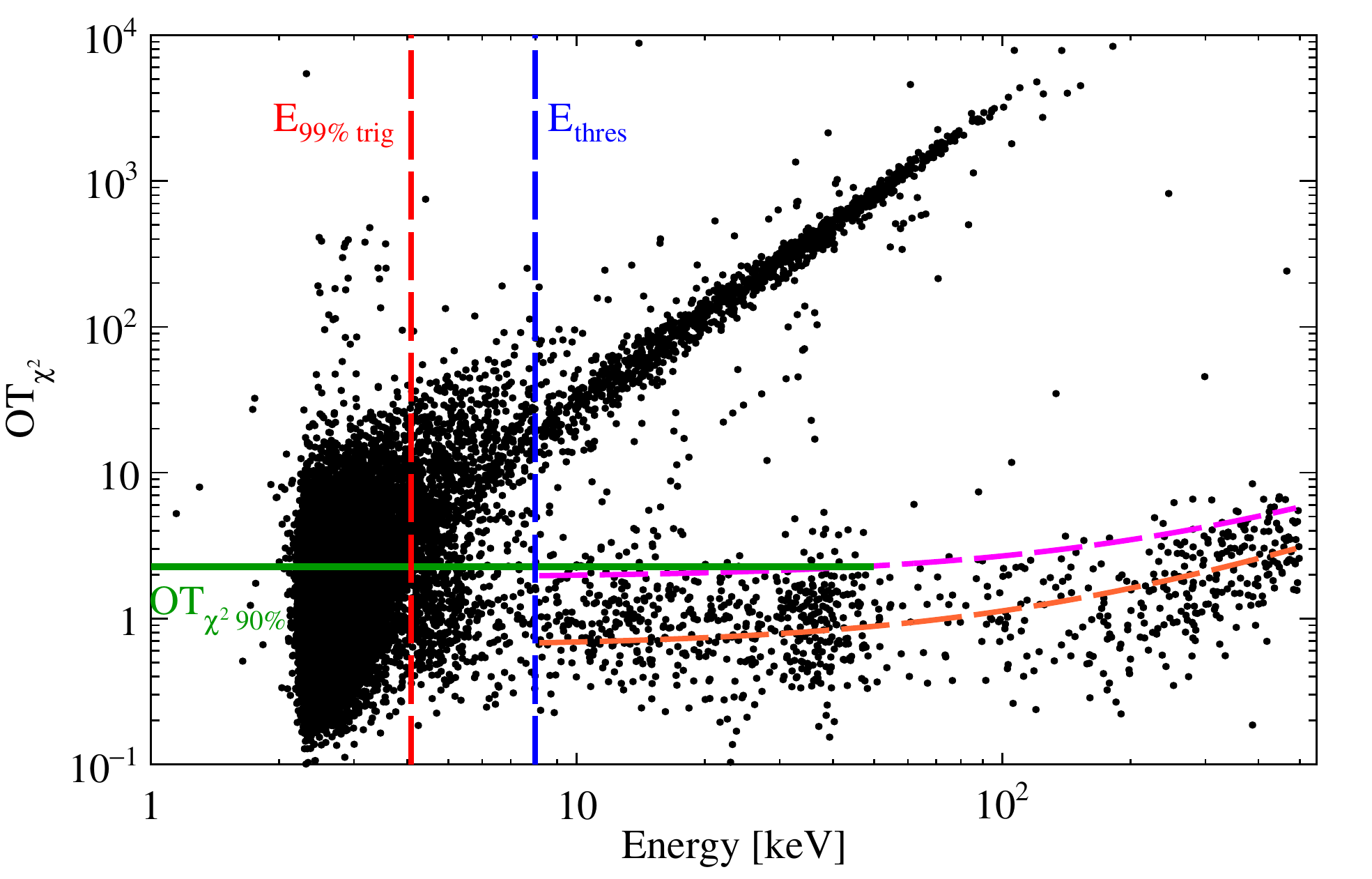}
\caption{(color online) A typical distribution of {\OT} as a function
  of energy for one BoDs. Physical events due to the particle
  interactions in the {\TeO} crystals are distributed in a
  nearly-horizontal band around {\OT} $\sim 1$. Non-physical events
  such as electronic noise and tower vibrations, as well as particle
  interactions in the thermistors, follow an oblique distribution. The
  green solid line corresponds to the 90th percentile of the {\OT}
  distribution ({\OTninty}) calculated in the region $[35-50]\1{keV}$
  and {\OT}$<10$ using physics data.  The magenta (orange) dashed line
  corresponds to the 90th (50th) percentile calculated using
  calibration data in the region $[100-500]\1{keV}$ and {\OT}$<10$,
  assuming linear dependence on energy. Red and blue dashed vertical
  lines represent the trigger threshold {\TT} and the analysis
  threshold $E_\n{thres}$, respectively.}
\label{fig:ot_chisquare}
\end{figure}

Fig.~\ref{fig:ot_chisquare} shows a typical {\OT} distribution as a
function of energy for the triggered events which pass the selection
criteria described in Sec.~\ref{subsec:ds_ch_selection_criteria}.
Between the OT trigger level and {\TT} there can be a leakage of baseline noise.
Physical events due to particle interactions in the {\TeO} crystals
scatter around {\OT}$\sim 1$ forming an almost horizontal
distribution, while spurious events due to electronic noise or 
particle interactions in the thermistors follow an oblique
distribution with {\OT} values as high as $\sim 10^4$ at $200\1{keV}$.
Pile-up events lie between the two bands.  We select legitimate
physical events with a requirement on the {\OT} parameter, and we
evaluate the selection efficiency by counting the number of events
before and after the cut in a region free of noise.  As is evident in
Fig.~\ref{fig:ot_chisquare}, {\OT} has a slight dependence on energy;
this dependence is more or less pronounced depending on the bolometer,
but mostly is imperceptible below $100\1{keV}$. Assuming no energy
dependence at low energy in the range between 10 and
$50\1{keV}$, we compute the selection efficiency in the region with
{\OT}$<10$ and energy in the range
{\OTChisquareValueObtainmentEnergyRegion}, where the statistics are
higher and the noise leakage is negligible.
We choose the values of the selection to achieve 90\% efficiency and
calculate it as the 90th percentile of the {\OT} distribution
({\OTninty}, green solid line in Fig.~\ref{fig:ot_chisquare}). The
selection efficiency is computed independently for every BoDs, and the
uncertainty, evaluated taking into account the statistical fluctuation
in counting for each BoDs, ranges from {\OTCutUncertaintyMin} to
{\OTCutUncertaintyMax}.
%
%
We have investigated the {\OT} dependence on energy using calibration 
data, as the low statistics above $60\1{keV}$ make the results significantly 
uncertain in background data. The behaviour is well described by a linear 
fit up to $500\1{keV}$ and the {\OTninty} value at $35-50\1{keV}$ agrees 
with that calculated in background data for bolometers featuring low rate during calibration, 
like the one on Fig. 2, where the magenta dashed line corresponds to 90\% efficiency 
and the orange dashed line to 50\% efficiency. However, in general the larger 
pile-up probability during calibration runs shifts the {\OTninty} selection 
upwards with respect to the value in background runs, 
so in the following we use the value calculated in background and assume 
no energy dependence down to threshold. In order to estimate the uncertainty related 
to the choice of an energy independent cut efficiency, 
we have calculated the counting rate below $35\1{keV}$ after the cut 
assuming the same energy dependence measured in calibration, 
being the difference with respect to the energy-independent selection lower than the statistical error.



\begin{figure}[htbp]
\centering
\includegraphics[width=0.5\textwidth]{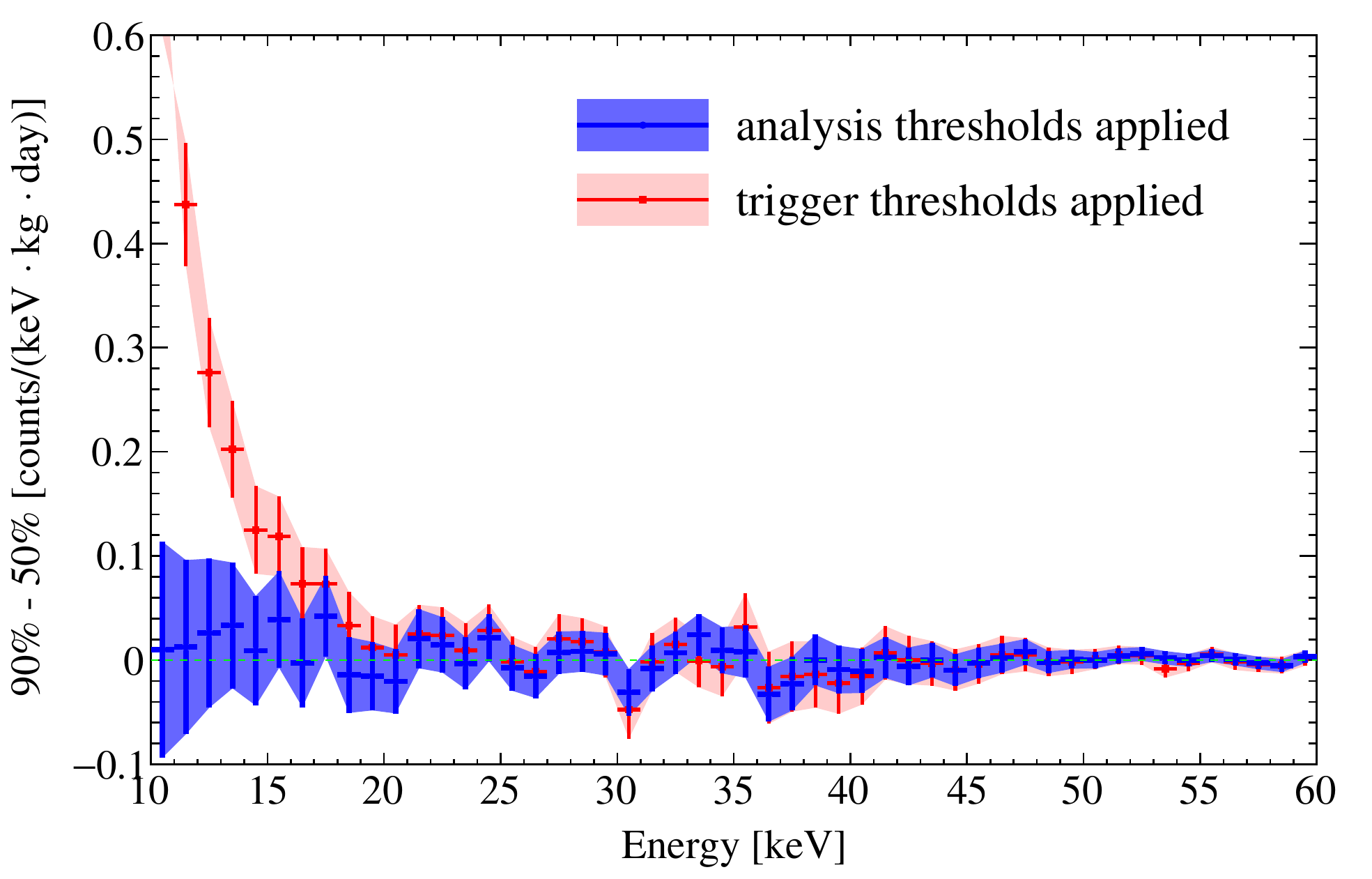}
\caption{(color online) {\OT} selection efficiency corrected rate
  difference between the background spectra calculated for {\OTninty}
  and {\OTfifty}. (See Sec.~\ref{subsec:threshold} for a description
  of analysis and trigger thresholds.)}
\label{fig:90_50_percentiles}
\end{figure}

The validity of the selection efficiency computation relies on the
assumption that the region {\OTChisquareValueObtainmentEnergyRegion}
is free of noise and the shape of the {\OT} distribution does not
change at lower energies. To verify this hypothesis, we compare the
{\OTninty} selection with the 50th percentile selection ({\OTfifty}),
assuming their selection efficiencies are 90\% and 50\%, respectively.
If there exists a significant noise contribution with the {\OTninty}
selection, the efficiency-corrected spectra would differ, as noise
rejection is stronger for the {\OTfifty} selection.  The residual
spectrum is shown in red in Fig.~\ref{fig:90_50_percentiles}. The
selection efficiency corrected rate difference between the two spectra
is compatible with zero down to {\EffCorRateCompZeroEne}. Below
{\EffCorRateCompZeroEne} the rate difference increases, suggesting the
presence of noise in the data.  In fact, the noise contribution for
most bolometers overlaps the physical-events band at energies directly above
{\TT}. In order to avoid noise contribution in the spectrum we set the most stringent
analysis energy threshold $E_\n{thres}>${\TT}, independently for every
BoDs, as described in Sec.~\ref{subsec:threshold}.

\subsection{Energy threshold determination}
\label{subsec:threshold}
The {\qz} cryostat at LNGS was more noisy than the R\&D cryostat in
which the CCVR bolometer performance tests were performed. It means
that in {\qz}, at $E=$~{\TT} we mainly trigger noise events. The pulse
shape parameter {\OT} presented in
Sec.~\ref{subsec:event_selection_criteria} provides powerful
discrimination between physics and spurious events,
but the two populations overlap as the energy decreases.

In order to avoid a leakage of spurious events in the data we set an
analysis threshold ({\Ethr}) at the minimum energy where the
populations are well separated for each BoDs.  Specifically, we perform
a Kolmogorov-Smirnov (KS) test to quantify the similarity between the
{\OT} populations in different energy slices with respect to the {\OT}
distribution at {\OTChisquareValueObtainmentEnergyRegion} and
{\OT}$<$10, the same pure signal sample region used to calculate the
{\OT} selection efficiency.  Starting from {\TT} (the vertical red
dashed line in Fig.~\ref{fig:ot_chisquare}), we compare the
distribution of {\OT} in $4\1{keV}$-width energy windows with the
reference distribution, for {\OT}$<10$. We set {\Ethr} at the lower
edge of the energy range that provides KS probability larger than
{\KSProbThr}.  This method is not valid when the reference sample is
contaminated with noise. We ensure the validity of the reference by
requiring {\OTninty}$< 6$. The value 6 is obtained from the {\OTninty}
distribution of all BoDs as the 2$\sigma$ above the median. We
discard {\NumberOfDsBoPairsDiscardedByOTLarger} BoDs that do not
fulfill this requirement from a total of {\NumberOfDsBoPairs}.
Once the KS threshold was fixed, the technique described worked for all 
{\NumberOfDsBoPairs} BoDs without manual adjustments, 
making it suitable for an $\mathcal{O}(1000)$ bolometers experiment.

Final analysis thresholds range from {\FinalEnergyThresholdMin} to
{\FinalEnergyThresholdMax}, with only {\NumberOfDsBoPairsWithThLower}
BoDs having a threshold lower than
{\FinalEnergyThresholdMinInSpc}, thus not being representative of the whole 
data-taking. Therefore, we set
{\FinalEnergyThresholdMinInSpc} as the minimum {\qz} energy threshold. The
exposure ranges from {\SelectedThExposureAtThreshold} at
{\FinalEnergyThresholdMinInSpc} up to {\SelectedThExposure} at
{\FinalEnergyThresholdMax} (see inset of Fig.~\ref{fig:le_spectrum}).
We verify that the noise acceptance is negligible with the same
procedure used in Sec.~\ref{subsec:event_selection_criteria}.  As
shown as the blue band in Fig.~\ref{fig:90_50_percentiles}, the
90\%-50\% residual with the analysis thresholds is compatible with
zero down to {\FinalEnergyThresholdMinInSpc}.

\subsection{Anti-coincidence requirement}
\label{sec:coincidence}
The last event selection criterion for the low energy rare event
searches is anti-coincidence; i.e., no signal events are triggered in
other bolometers in a certain temporal window.  We use a coincidence
window of {\LECoincidenceTime}, 20 times wider than that used for the
standard {\BBless} analysis~\cite{Alduino:2016zrl}, due to the larger
difference in characteristic rise time between low and high energy
events.  
To evaluate the event loss due to random coincidences between physical 
events and unrelated events on another bolometer (anticoincidence selection efficiency)
we use the $1461\1{keV}$ $\gamma$ ray peak in the single crystal energy spectrum. 
This peak, coming from {\K} EC, does not belong to any cascade, so the only true
coincident event is the $\sim 3\1{keV}$ X-ray from the Ar de-excitation, which is below our threshold.
Counting the number of
events in the $1461\1{keV}$ peak of the single crystal spectrum before and after 
the selection we find the anti-coincidence selection efficiency to be
{\AntiCoincidenceEff}. We combine this efficiency with the 90\% event
selection efficiency on {\OT} to obtain the total detection
efficiency. The uncertainty on the efficiency is BoDs dependent.

\section{Low energy spectrum construction}
\label{sec:low_energy_spectrum}

\subsection{Energy calibration}
\label{subsec:energy_calibration}
During the calibration runs, the ${}^{232}\n{Th}$ sources are outside
the cryostat, so the $\gamma$ rays pass through a
$\mbox{1.4-cm-thick}$ ancient Roman lead shield before reaching the
detector.  Consequently, the peaks in the low energy region of the
spectrum are highly attenuated, and only those between 511 and
$2615\1{keV}$ are clearly visible and used to calibrate the energy
response of each bolometer.  As stated in Sec.~\ref{sec:exp_setup}, in
the standard {\BBless} analysis we use a second-order polynomial with
zero intercept to fit the reconstructed peak positions in the
calibration spectrum to their nominal energies. To improve the 
energy resolution at the $Q$-value of {\BBless}, a combination of four
energy estimators with slightly different calibration coefficients are
used in the final analysis presented in~\cite{Alfonso:2015wka}. For
the low energy analysis, we use a single energy estimator to 
avoid complexity. 

The calibration uncertainty in the {\BBless} ROI is
{\StandCalibUncertaintyAtROI}~\cite{Alduino:2016zrl}, but this value
is energy dependent.  To validate the extrapolation to energies below
$100\1{keV}$, we use the characteristic Te X-rays that can follow a
$\gamma$ ray interaction, which can escape the crystal and be detected
in another crystal.  These events can be selected by requiring
a coincidence in an adjacent crystal. The most intense X-rays are
eight K-shell peaks ranging from {\TeXrayPeakRangeMin} to
{\TeXrayPeakRangeMax} (see Table~\ref{tab:xrays}). Due to the energy resolution, these
peaks are noticeable as a main and a secondary peak around
{\TeXrayFirstPeakPos} and {\TeXraySecondPeakPos}, in both the
calibration and background spectra.

\begin{table}[ht]
\begin{center}
\begin{tabular}{@{\extracolsep{\fill}} c c c c c}
\hline
\hline
Line & Energy (keV) & Intensity (\%) \\
\hline
K$\alpha$1  &  27.472  &   47.1     \\
K$\alpha$2  &  27.202  &   25.3 \\
K$\alpha$3  &  26.875  &  0.00202 \\
\hline
K$\beta$1   &  30.995  &   8.19 \\
K$\beta$2   &  31.704  &   2.37 \\
K$\beta$3   &  30.944  &   4.25 \\
K$\beta$4   &  31.774  &   0.363   \\
K$\beta$5   &  31.237  &   0.075 \\
\hline
\hline
\end{tabular}
\end{center}
\caption{Main Te X-ray emission lines. Data from~\cite{nndc}.}
\label{tab:xrays}
\end{table}

%
To determine their reconstructed energies, we fit the region from
{\CalibrationPeakFitRangeMin} to {\CalibrationPeakFitRangeMax} with an
eight-Gaussian line shape plus a linear background, where all of the
Gaussians are constrained to have the same width. The relative
intensities and positions of each Gaussian are fixed with respect to
the main $\n{K}_{\alpha \n{1}}$ peak using nuclear data
from~\cite{nndc}.  
In order to take into account any possible
discrepancy in the relative intensities of the peaks arising from
systematic effects in the detector, we determine these intensities
with a Monte Carlo (MC) simulation based on the 
Geant4 package~\cite{geant4} (version 4.9.6.p03, see ~\cite{Alduino:2017qet} for details)
that includes the bolometer-dependent energy thresholds and the analysis coincidence
window. 

\begin{figure}[ht]
\centering 
\centering \includegraphics[width=0.5\textwidth]{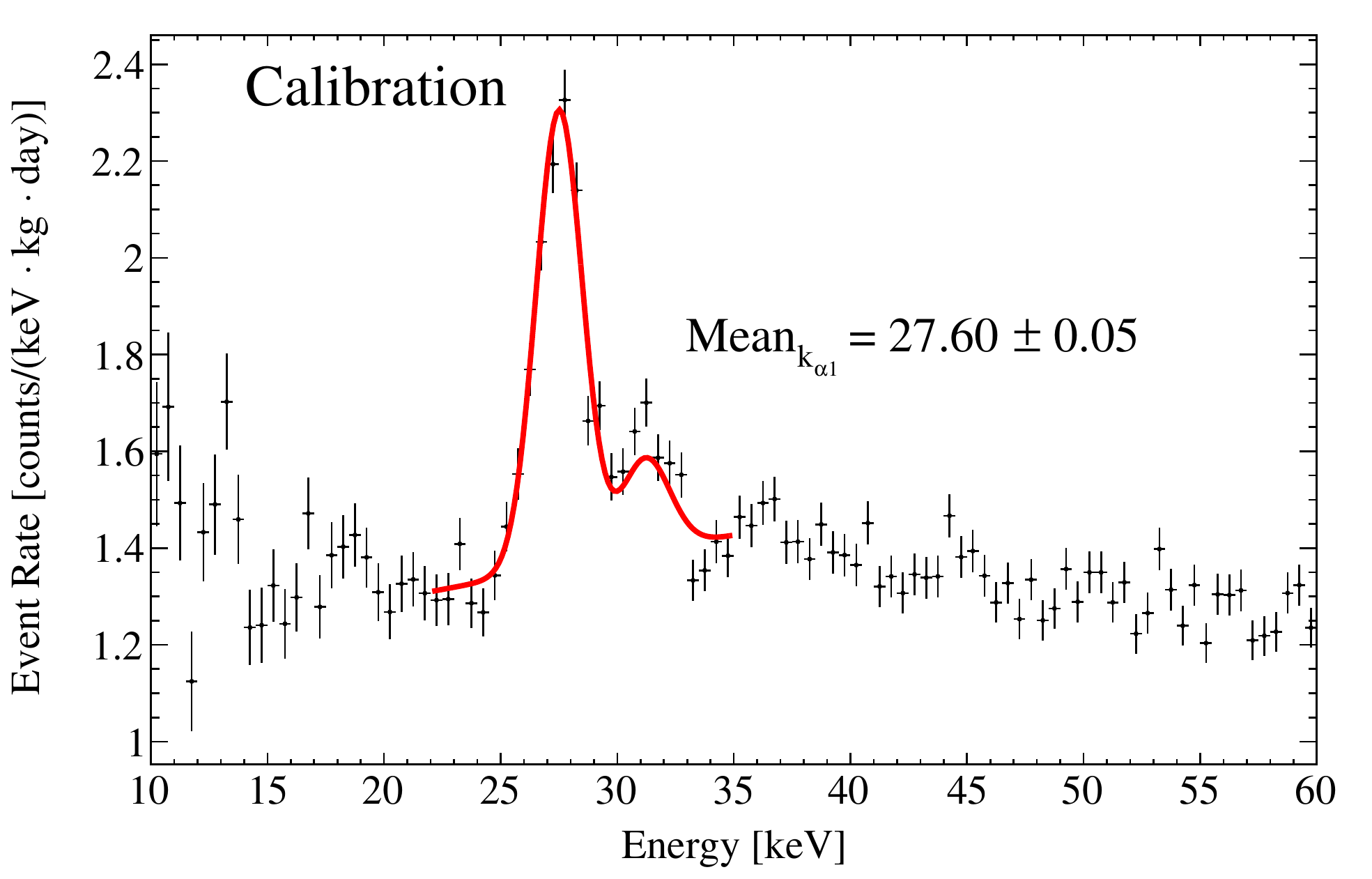}
\centering \includegraphics[width=0.5\textwidth]{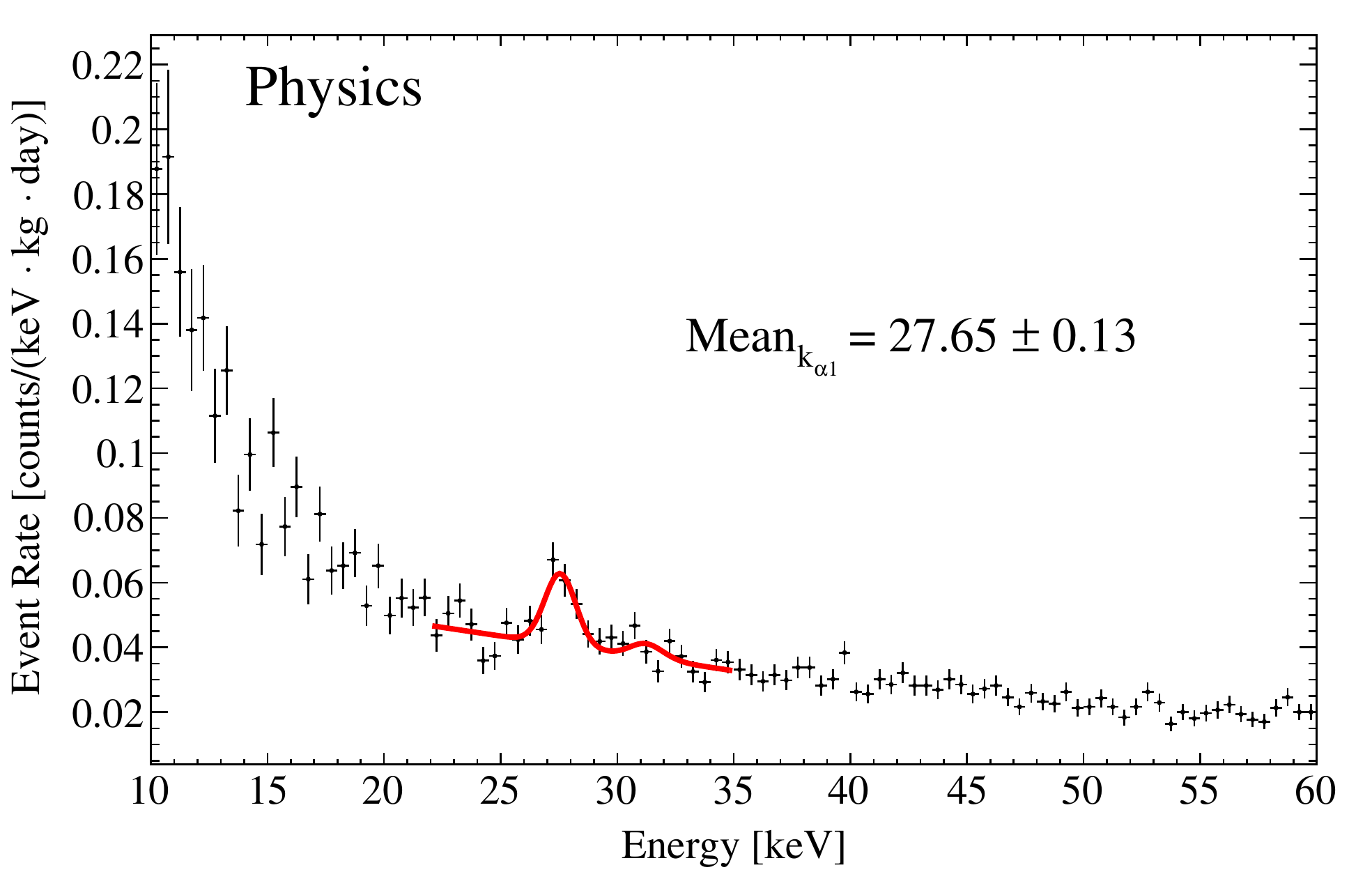}
\caption{(color online) {\qz} summed energy spectra of events with
  double-crystal coincidence in calibration (top) and physics (bottom)
  data, along with fits to an eight-Gaussian line shape plus linear
  background (red solid lines).}
\label{fig:calbration_peaks_fits}
\end{figure}

Fig.~\ref{fig:calbration_peaks_fits} displays the fit results of both
calibration and physics data.  The most intense $\n{K}_{\alpha\n{1}}$
peaks for calibration and physics data are measured to be
{\MostIntenseLowEPeakCal} and {\MostIntenseLowEPeakBkg},
respectively. The corresponding residuals with respect to the nominal
energy are {\MostIntenseLowEPeakDiffCal} and
{\MostIntenseLowEPeakDiffBkg}, respectively. The latter indicates the
systematic upward shift on the energy scale in the physics data, and we
take into account its impact on the WIMP sensitivity reported in
Sec.~\ref{sec:wimp_analysis}.

The difference in peak positions between the nuclear data and
the simulation is found to be less than {\MCPeakPositionDiff}.
Nevertheless, the amplitude ratio $\n{K}_{\beta \n{1}}/\n{K}_{\alpha \n{1}}$ 
in the MC simulation is 0.27 instead of 0.17 from Table~\ref{tab:xrays} due to the strong change in X-ray
attenuation length between $27$ and $31\1{keV}$. 
This effect is also appreciable in the CUORE-0 data; as shown in 
Fig.~\ref{fig:calbration_peaks_fits}
(upper panel) the $31\1{keV}$ peak is underestimated by the fit function. 
Leaving the relative intensity $\n{K}_{\beta \n{1}}$/$\n{K}_{\alpha \n{1}}$ 
as a floating parameter improves the goodness of the fit and reproduces 
the relative intensities estimated by MC, but the position of the $\n{K}_{\alpha \n{1}}$ 
peak does not change within the uncertainty.



\subsection{Energy spectrum}

\begin{figure}[ht]
\centering
\includegraphics[width=0.5\textwidth]{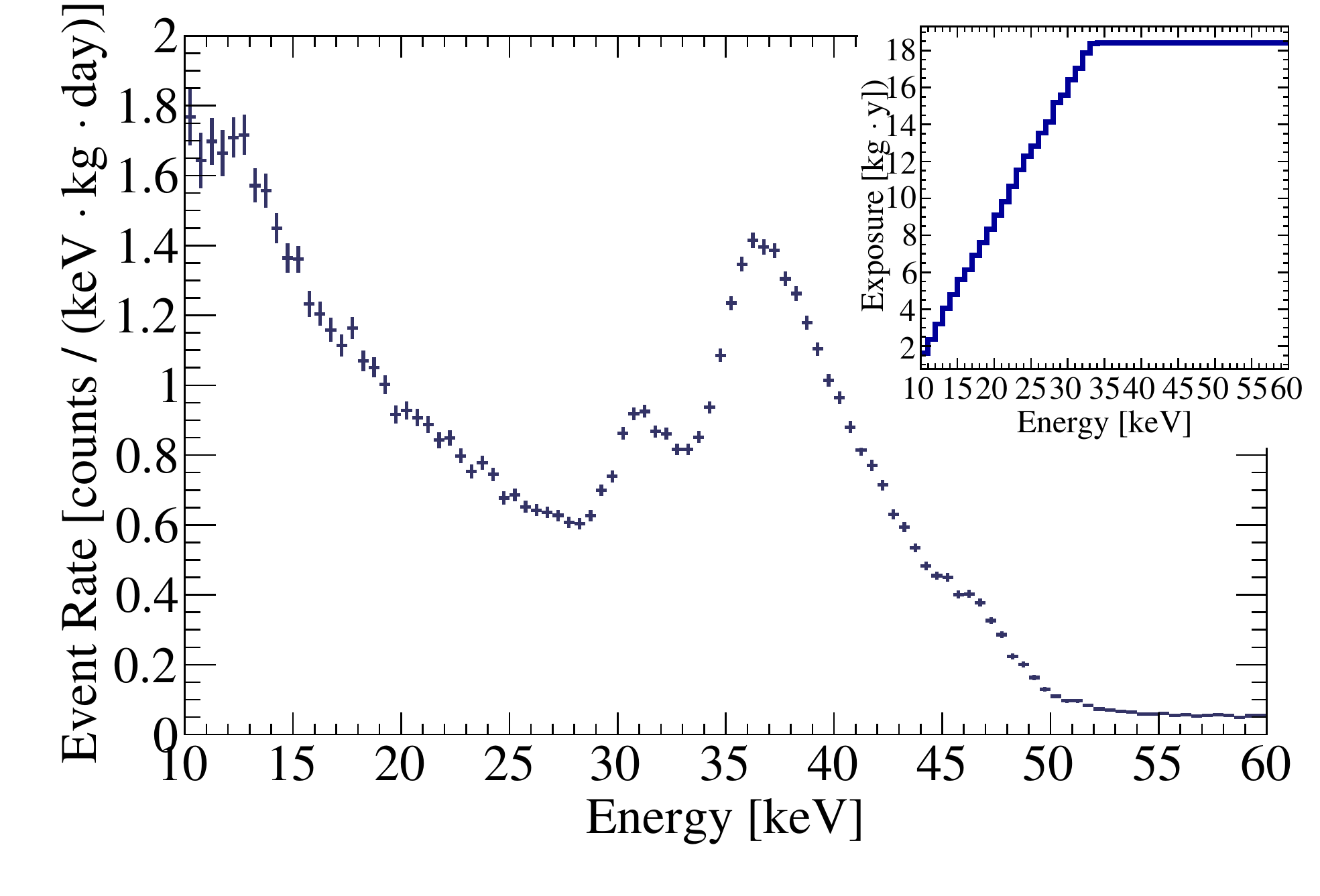}
\caption{Efficiency-corrected energy
  spectrum of {\qz} from 10 to $60\1{keV}$ corresponding to an exposure
  ranging from {\SelectedThExposureAtThreshold} at
  {\FinalEnergyThresholdMinInSpc} up to {\SelectedThExposure} at
  $35\1{keV}$, as shown in the inset.
}
\label{fig:le_spectrum}
\end{figure}

Fig.~\ref{fig:le_spectrum} shows the low energy spectrum of {\qz},
using the selected BoDs with the event selection criteria
described in Sec.~\ref{sec:data_selection} and the detection
efficiency. The background rate above $50\1{keV}$ is {\BkgAtFiftykeV},
consistent with the results obtained in the standard {\BBless}
analysis. Below $50\1{keV}$, the background rate increases substantially
to {\BkgAtTenkeV} at $10\1{keV}$; it is, however, two times lower
than the background rate measured with the four best {\qino}
bolometers for which thresholds below $10\1{keV}$ were
attained~\cite{Alessandria:2012ha}.

The most noticeable feature in the spectrum is a peak-like structure
around {\LowEPeakLow} and {\LowEPeakHigh}. Given that this structure
is observed in all the bolometers and it was also present in the Cuoricino background, 
its origin is likely physical.
The current background prediction based on our MC simulation~\cite{Alduino:2017qet} does not fully
account for the low energy spectrum including this peak-like
structure, and further investigation is on-going
under the hypothesis that the contamination is due to materials facing the detectors (e.g. copper shielding...).
We expect to have
better insights on this peak-like structure with CUORE data, where
inner bolometers mostly face other bolometers and not the copper
shielding. Through the comparison between the innermost and outermost
bolometers, we may be able to attribute the origin of this peak-like
structure to a certain process.


\section{CUORE sensitivity to WIMP annual modulation}
\label{sec:wimp_analysis}
In this section we present the sensitivity of CUORE 
to the annual modulation in the detection rate induced by 
dark matter in the galactic halo. We restrict our analysis 
to WIMPs interacting with the target nuclei in the detector 
via elastic scattering off nuclei; for this reason, we
present a study of the nuclear quenching factor of {\TeO} obtained in {\qz}.

\subsection{{\TeO} nuclear quenching factor}
One of the prerequisites to perform a WIMP dark matter search is a
good understanding on the low energy response of the detector for both
nuclear recoils (NRs), produced by WIMPs or background neutrons, and
electronic recoils (ERs), produced by electromagnetic backgrounds.
The nuclear quenching factor is defined as the ratio of the measured
signal generated by a NR to that generated by an ER depositing the
same energy in the detector, and depends on the energy and recoiling
nucleus. Given that any energy conversion in the {\TeO} bolometers finally
produces signal through phonons, the nuclear quenching factor in the
bolometers is expected to be close to one.
The nuclear quenching factors of several recoiling nuclei in {\TeO}
have been measured previously using the daughter nuclei of the
$\alpha$ decays from ${}^{224}\n{Ra}$, ${}^{220}\n{Rn}$,
${}^{216}\n{Po}$, ${}^{212}\n{Po}$, and ${}^{212}\n{Bi}$ in the energy
range between 100 and $170\1{keV}$. The result was found to be
$1.025\pm0.01\,{\rm (stat)}\pm0.02\,{\rm (syst)}$~\cite{Alessandrello:1997ca}. We exploit the same
technique and estimate the nuclear quenching factors 
using the daughter nuclei of some $\alpha$ emitters 
at energies around $100\1{keV}$.

Specifically, we measure the recoiling energy of the daughter nuclei
following $\alpha$ decays of ${}^{210}\n{Po}$, ${}^{222}\n{Rn}$,
${}^{224}\n{Ra}$, and ${}^{218}\n{Po}$ present in the {\qz} crystal
surfaces where either the $\alpha$ particle or the daughter nucleus
escapes and is detected in an adjacent crystal.  We tag these events by
requiring coincidence in two crystals with a total energy within some
tens of keV of the $Q$-value of the decay.  Then we fit the spectrum
of the recoiling nuclei with an
asymmetrical Gaussian function with a smooth power-law tail
relative to the mean to obtain the peak position.
Table~\ref{tab:measured_qf} summarizes the obtained nuclear quenching
factors for the selected recoiling nuclei. While the nuclear quenching
factor obtained from ${}^{218}\n{Po}$ (${}^{214}\n{Pb}$ and
${}^{220}\n{Rn}$) is close to unity, we notice that
the one obtained from
${}^{206}\n{Pb}$ exhibits significant deviation from unity. 


%
\begin{table}[htbp]
  \centering
  \caption{Expected energy, measured energy obtained from fits,
    and resulting quenching factors (QFs) for the selected recoiling
    nuclei. Only statistical uncertainties are shown in the QFs.}
  \label{tab:measured_qf}
  \begin{tabular*}{1.0\columnwidth}{@{\extracolsep{\fill}} c c c c c}
    \hline\hline Recoiling Nuclei & $E_\n{expected}$ & $E_\n{measured}$ & QF\\
    \hline
    ${}^{206}\n{Pb}$ & 103.12 & $95.62 \pm 0.24$  &  $0.927 \pm 0.002\n{(stat.)}$ \\
    ${}^{218}\n{Po}$ & 100.8 & $100.0 \pm 0.9$ & $0.992 \pm 0.009\n{(stat.)}$ \\
    ${}^{220}\n{Rn}$ & 103.50 & $100.45 \pm 1.21$ & $0.971 \pm 0.012\n{(stat.)}$  \\
    ${}^{214}\n{Pb}$ & 112.13 & $110.92 \pm 0.96$  & $0.989 \pm 0.009\n{(stat.)}$\\
    \hline \hline
  \end{tabular*}
\end{table}

Acknowledging that these recoiling nuclei are surface events and
energy losses might happen at the surface, we use unity as the nuclear
quenching factor to set the energy scale of WIMPs in the following
analysis, and conservatively estimate an uncertainty of nuclear
quenching factor as {\NRQuenchingUncertainty} using the largest
deviation from unity observed by ${}^{206}\n{Pb}$. Its impact is
integrated in Fig.~\ref{fig:wimp_sensitivity} to report the WIMP
sensitivity of CUORE.


\subsection{WIMP sensitivity of CUORE}
The sensitivity of {\qz} to annual modulation of WIMPs is limited by
its relatively small exposure.  However, we can use results of {\qz}
to estimate the CUORE sensitivity assuming the same background rate
and energy thresholds. This is a conservative hypothesis since
a significant background reduction is
expected to be achieved in CUORE thanks to its close-packed detector
configuration and the careful selection of radio-pure detector materials~\cite{Alduino:2017qet}.

{\TeO} is an interesting DM target, as it combines a heavy nucleus (tellurium), 
which provides a large scattering amplitude (assuming coherent interaction, that scales as A$^2$) 
and a light one (oxygen) to enhance sensitivity in the low-mass WIMP region.
To calculate the expected WIMP rate in the detectors, we follow the
commonly used analysis framework for WIMP direct
detection~\cite{Smith:1988kw,Lewin:1995rx}. 
We consider only the
spin-independent (SI) contribution since the spin-dependent
contribution is comparatively reduced in {\TeO}; the main isotopes
with non-zero nuclear spin are ${}^{125}\n{Te}$ and ${}^{123}\n{Te}$,
with isotopic abundance of $7.1\%$ and $0.9\%$, respectively. We assume
coherent isospin-invariant coupling and the Helm model~\cite{Helm:1956zz} for
the nuclear form factors.  Under these assumptions, the generic WIMP
is completely determined by its mass {\mw} and SI WIMP-nucleon cross
section {\ssi}. For the velocity distribution of dark matter, we use
the standard halo model (SHM)~\cite{Freese:2012xd} commonly adopted for comparisons of direct
detection experiments. Consequently, the annually modulating WIMP
recoil rate due to the motion of the Earth around the Sun can be
approximated using a constant term $S_0$ plus a cosine-modulated term
$S_m$, as given by

\begin{equation}
\frac{dR_W}{dE}(E,t) = S_{0}(E) + S_{m}(E) \n{cos}[\omega (t-t_0)]
\label{eq:rateMod}
\end{equation}
where $\omega=2\pi/\n{yr}$ and $t_0$ is around June 1.

To obtain the sensitivity to annual modulation of WIMPs, we scan over
the WIMP parameter space ({\mw}, {\ssi}) looking for the region at
which a WIMP interaction would produce an annual modulation in the
detection rate over the measured background at a certain confidence
level (C.L.). For each ({\mw}, {\ssi}) we generate 100 toy MC
simulations and for each MC spectrum, we perform a maximum likelihood
(ML) analysis for both the annual-modulation (AM) and the absence of
modulation (null) hypotheses. We quote the significance of the
modulation as the log-likelihood ratio of the best fits
$\chi^2$=2log($\mathcal{L}_\n{AM}/\mathcal{L}_\n{null}$).  The
likelihood $\mathcal{L}_\n{AM}$ is calculated using the probability
density function (PDF)
\begin{align}
\phi = \frac{dR_W}{dE}(E,t;m_W,\sigma_\n{SI})M_\n{det}\epsilon_\n{BoDs}(E,t)
+\phi_\n{b}(E; a_i)\epsilon_\n{BoDs}(E,t)
\end{align}
where $M_\n{det}$ is the target mass, $\epsilon_\n{BoDs}(E,t)$ is the
BoDs-dependent detection efficiency, and $\phi_\n{b}$ is the
background PDF, which we model with a Chebychev polynomial with
coefficients $a_i$  and for which we do not consider any temporal dependence. 
The likelihood $\mathcal{L}_\n{null}$ is
calculated from $\phi_\n{b}$ alone.

We choose the ROI to perform the analysis as
{\WIMPAnalysisEnergyRange}, which excludes the peak-like structures
above $30\1{keV}$ shown in Fig.~\ref{fig:le_spectrum}. Given that the
differential rate of WIMPs quasi-exponentially falls as a function of
energy, most of the signal is contained at the low energy and the
expected contribution to the WIMP sensitivity from $30-60\1{keV}$
is negligible compared to that from {\WIMPAnalysisEnergyRange}.
We consider a target mass of {\CUORETotalMass} and the scheduled 5
years of data-taking with 75\% duty cycle, accounting for the
calibration time. Based on the {\qz} energy
threshold, we use $10\1{keV}$ but we also show the sensitivity that
could be attained under the more optimistic hypothesis that we reach a
$3\1{keV}$ threshold as demonstrated in the CCVR experiment with a
linear extrapolation of the {\qz} background to lower energies.

\begin{figure}[htp]
\centering 
\includegraphics[width=0.5\textwidth]{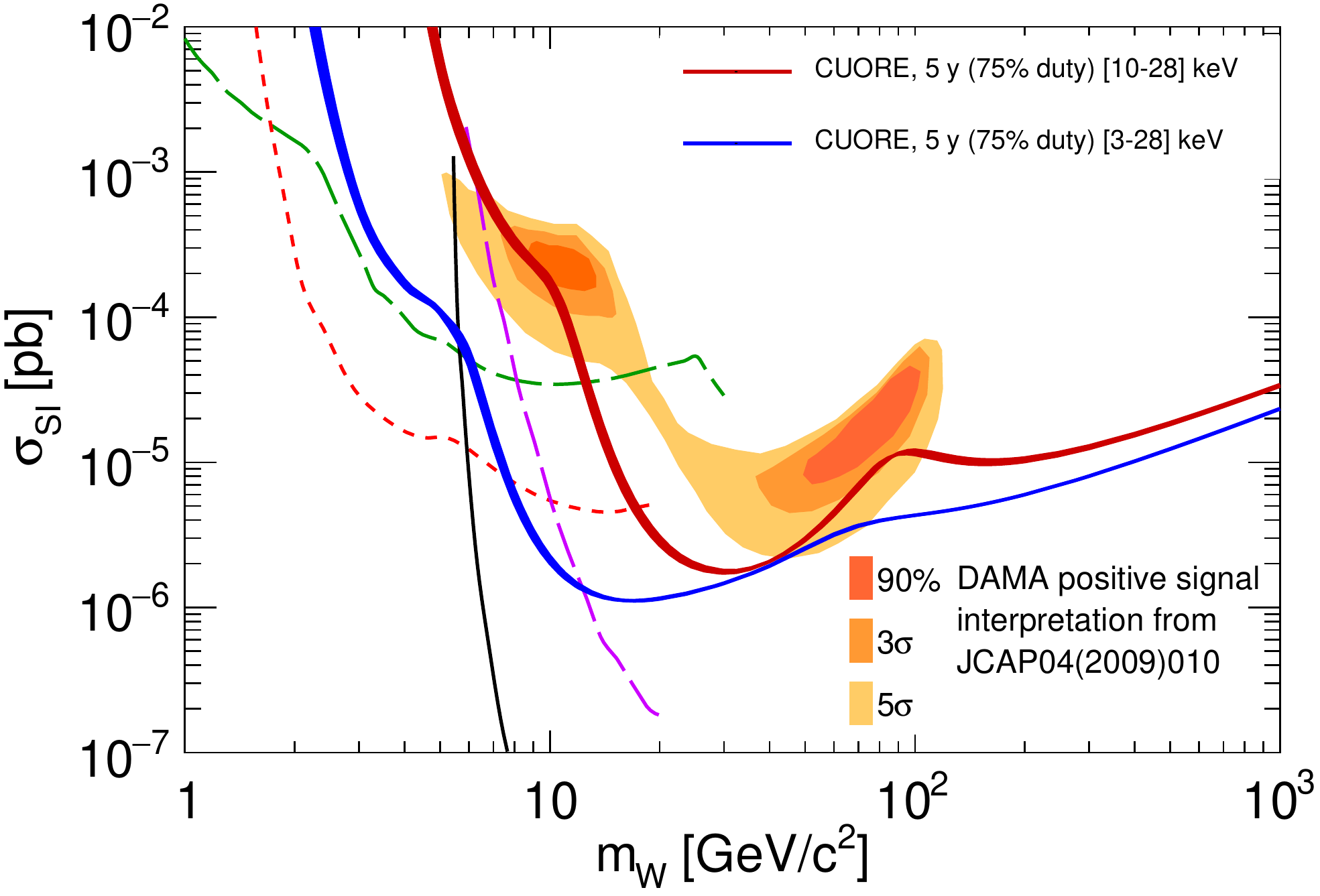}
\caption{(color online) 90\% sensitivities on the SI elastic
  WIMP-nucleon cross section as a function of WIMP mass of CUORE,
  assuming 5 years of data-taking with 75\% of duty cycle and
  $10\1{keV}$ threshold (red), as well as $3\1{keV}$ threshold
  (blue). Uncertainty on the energy scale dominated by the nuclear
  quenching factor is taken into account. DAMA/LIBRA positive signal
  reported in~\cite{Savage:2008er} is shown as yellow/dark
  yellow/orange islands. The results from CRESST-II (dashed
  green)~\cite{Angloher:2015ewa}, CDMS Lite (dashed
  red)~\cite{Agnese:2015nto}, XMASS (dashed
  violet)~\cite{Abe:2015eos}, and LUX (black
  solid)~\cite{Akerib:2013tjd} are also shown.}
\label{fig:wimp_sensitivity}
\end{figure}

Fig.~\ref{fig:wimp_sensitivity} shows CUORE sensitivity requiring a
90\% C.L. in 90\% of the toy-MC experiments. The results are
consistent with those obtained with a pure statistical calculation
following~\cite{Cebrian:1999qk}. This figure assumes a WIMP local
density {\WIMPLocalDensity}, local circular velocity
{\LocalCircularVelocity}, galactic escape velocity
{\GalacticEscapeVelocity}, and orbital velocity of the Earth around
the Sun {\EarthOrbitalVelocity}. Uncertainty on the energy scale
dominated by the nuclear quenching factor is taken into account.  For
comparison, we also show the 5$\sigma$, 3$\sigma$ and 90\%
C.L. regions resulting from a ML analysis reported
in~\cite{Savage:2008er} on the DAMA/LIBRA annual modulation positive
result~\cite{Bernabei:2008yi,Bernabei:2013xsa} using the same
parameters for the SHM.  Thanks to the $741\1{kg}$ of
target mass of CUORE, we expect to achieve the sensitivity required to
fully explore the parameter region implied by the DAMA/LIBRA positive
annual modulation signal with 5 years of data-taking. Other recent
experimental results from CRESST-II, CDMS Lite, XMASS and
LUX~\cite{Angloher:2015ewa,Agnese:2015nto,Abe:2015eos,Akerib:2013tjd}
are also shown. The results from CRESST-II, CDMS, and LUX were
obtained using $v_\n{esc} = 544\1{km/s}$. The impact of using
$v_\n{esc} = 544\1{km/s}$ instead of $v_\n{esc} = 650\1{km/s}$ for
CUORE sensitivity is found to be less than {\VescImpactUpLimit} at
$6\1{GeV}$.
Also for the other experiments only a minor impact of the escape velocity on the exclusion limit is expected.

\section{Summary}
\label{sec:summary}
We have presented the analysis techniques developed for low energy
rare event searches with CUORE and their validation using the data
acquired with the {\qz} experiment. 
We have optimized the software trigger developed in previous CUORE 
prototypes, removing an intrinsic dead time that prevented the algorithm 
from reaching 100\% efficiency, and designed a
protocol to periodically monitor the efficiency by injecting
low energy reference pulses at the end of every dataset. With the new
trigger, we have reduced the {\qz} trigger thresholds from several
tens of keV to values between 4 and $12\1{keV}$.

We have also demonstrated that a pulse shape analysis can efficiently
select legitimate physics events in {\TeO} bolometers
against spurious ones at energies below $100\1{keV}$.
In addition, we have developed a
technique, scalable to an experiment with one thousand bolometers, to
independently establish the analysis threshold of each bolometer in
each dataset. In {\qz} the analysis thresholds range between 8 and $35\1{keV}$.
Using characteristic X-rays from Te, we have found the
energy scale shift to be {\MostIntenseLowEPeakDiffBkg} upward at $\sim
27\1{keV}$.

After the data and event selection, the {\qz} 
background rate ranges from {\BkgAtTenkeV} at $10\1{keV}$ to
{\BkgAtFiftykeV} at $50\1{keV}$, two times less than that attained
with the best {\qino} bolometers.  Nevertheless, the low energy
spectrum requires further investigation including the explanation of
the peak-like structures between 30 and $40\1{keV}$. We use the
nuclear quenching factors of {\TeO} obtained by tagging the recoiling
daughter nuclei of $\alpha$ decays in the {\qz} data to estimate the
uncertainty of the WIMP energy scale. We incorporate it to report the
CUORE sensitivity to WIMP annual modulation.

CUORE will search for low energy rare events such as solar axions,
WIMP dark matter in the galactic halo, or coherent scattering of
galactic supernova neutrinos using the analysis techniques presented
in this paper.  In particular, we expect to reach a sensitivity to
annual modulation of WIMPs sufficient to fully explore the parameter
region indicated by the positive annual modulation signal of the
DAMA/LIBRA experiment with 5 years of CUORE data-taking.

\section*{Acknowledgments}
The CUORE Collaboration thanks the directors and staff of the Laboratori Nazionali del Gran Sasso and our technical staff for their valuable contribution to building and operating the detector. 
This work was supported by the Istituto Nazionale di
Fisica Nucleare (INFN); the National Science
Foundation under Grant Nos. NSF-PHY-0605119, NSF-PHY-0500337,
NSF-PHY-0855314, NSF-PHY-0902171, NSF-PHY-0969852, NSF-PHY-1307204, and NSF-PHY-1404205; the Alfred
P. Sloan Foundation; the University of Wisconsin Foundation; and Yale
University. This material is also based upon work supported  
by the US Department of Energy (DOE) Office of Science under Contract Nos. DE-AC02-05CH11231 and
DE-AC52-07NA27344; and by the DOE Office of Science, Office of Nuclear Physics under Contract Nos. DE-FG02-08ER41551 and DEFG03-00ER41138. 
This research used resources of the National Energy Research Scientific Computing Center (NERSC).

\bibliographystyle{spphys}
\bibliography{cuore_paper}

\end{document}